\documentclass[aps,prl,twocolumn,showpacs,floatfix]{revtex4-1}
\usepackage{graphicx,amsfonts,amssymb,amsmath,hyperref}
\usepackage{textcomp}

\newif\ifhyper
\hypertrue
\ifhyper
\hypersetup{
  citecolor = {green},
  colorlinks = {true}, 
  urlcolor = {blue} 
}

\begin{document}

\graphicspath{{./figures_submit/}}



\newcommand{\ket}[1]{|#1\rangle}
\newcommand{\const}{{\rm const}} 
\newcommand{\mean}[1]{\langle #1 \rangle}

\newcommand{\ie}{i.e. }
\newcommand{\eg}{e.g. }
\newcommand{\cc}{{\rm c.c.}} 
\newcommand{\hc}{{\rm h.c.}} 

\def\eps{\epsilon}
\def\gam{\gamma} 
\def\lamb{\lambda}
\def\sig{\sigma}

\def\half{\frac{1}{2}}

\def\q{{\bf q}}
\def\r{{\bf r}}
\def\nablabf{\boldsymbol{\nabla}}

\def\w{\omega}
\def\wn{\omega_n}
\def\wnu{\omega_\nu}
\def\wp{\omega_p} 
\def\dmu{{\partial_\mu}}
\def\dl{{\partial_l}}  
\def\dt{\partial_t} 
\def\dtau{{\partial_\tau}} 

\def\intr{\int d^dr}  
\def\inttau{\int_0^\beta d\tau}

\title{Quantum criticality of a Bose gas in an optical lattice near the Mott transition} 

\author{A. Ran\c{c}on and  N. Dupuis}
\affiliation{Laboratoire de Physique Th\'eorique de la Mati\`ere Condens\'ee, 
CNRS UMR 7600, \\ Universit\'e Pierre et Marie Curie, 4 Place Jussieu, 
75252 Paris Cedex 05,  France}

\date{September 13, 2011}
 
\begin{abstract} 
We derive the equation of state of bosons in an optical lattice in the framework of the Bose-Hubbard model. Near the density-driven Mott transition, the expression of the pressure $P(\mu,T)$ versus chemical potential and temperature is similar to that of a dilute Bose gas but with renormalized mass $m^*$ and scattering length $a^*$. $m^*$ is the mass of the elementary excitations at the quantum critical point governing the transition from the superfluid phase to the Mott insulating phase, while $a^*$ is related to their effective interaction at low energy. We use a nonperturbative renormalization-group approach to compute these parameters as a function of the ratio $t/U$ between hopping amplitude and on-site repulsion. 
\end{abstract}
\maketitle

{\it Introduction.} 
Quantum phase transitions play a crucial role in many strongly-correlated systems ranging from quantum antiferromagnets to heavy-fermion materials and high-$T_c$ superconductors~\cite{[{See, for instance, }]Sachdev08,*Gegenwart08,*Sachdev10}. Although these transitions occur at zero temperature, theory predicts that the finite-temperature thermodynamics in the vicinity of a quantum critical point (QCP) is described by universal scaling relationships up to rather high temperatures~\cite{[{For a general introduction to quantum phase transitions, see }]Sondhi97,*Coleman05,*Sachdev11,Sachdev_book}. While understanding quantum criticality in strongly-correlated systems is often a challenge, both experimentally and theoretically, cold atoms offer clean systems for a quantitative and precise study of quantum phase transitions. Quantum criticality in cold atoms has attracted increasing theoretical interest in the last years~\cite{Zhou10,*Hazzard11,*Fang11}, and the experimental observation of the quantum critical behavior of a two-dimensional Bose gas in an optical lattice near the vacuum-superfluid transition has recently been reported~\cite{Zhang11}.  

In ultracold gases, strong correlations can be achieved by tuning the atom-atom interactions by means of a Feshbach resonance, or by loading the atoms into an optical lattice~\cite{[{See, for instance, }] Bloch08}. In the latter case, by varying the strength of the lattice potential, it is possible to induce a quantum phase transition between superfluid and Mott insulating ground states in a Bose gas~\cite{Jaksch98,*Greiner02}. The main features of this transition can be understood in the framework of the Bose-Hubbard model, which describes bosons hopping on a lattice with an on-site repulsive interaction~\cite{Fisher89}. 

The density-driven Mott transition belongs to the same universality class as the transition between the vacuum and the superfluid phase in a dilute Bose gas: both transitions are governed by the same QCP~\cite{Fisher89,Sachdev_book}. It follows that the pressure must take the same (universal) form near the QCP (up to some nonuniversal parameters as will be explained in detail below). In this Letter, we
prove this result by computing the pressure $P(\mu,T)$ (as a function of chemical potential and temperature) in various limits using a nonperturbative renormalization-group approach to the Bose-Hubbard model~\cite{Rancon11a,Rancon11b}. 

\begin{figure}
\centerline{\includegraphics[width=6cm,clip]{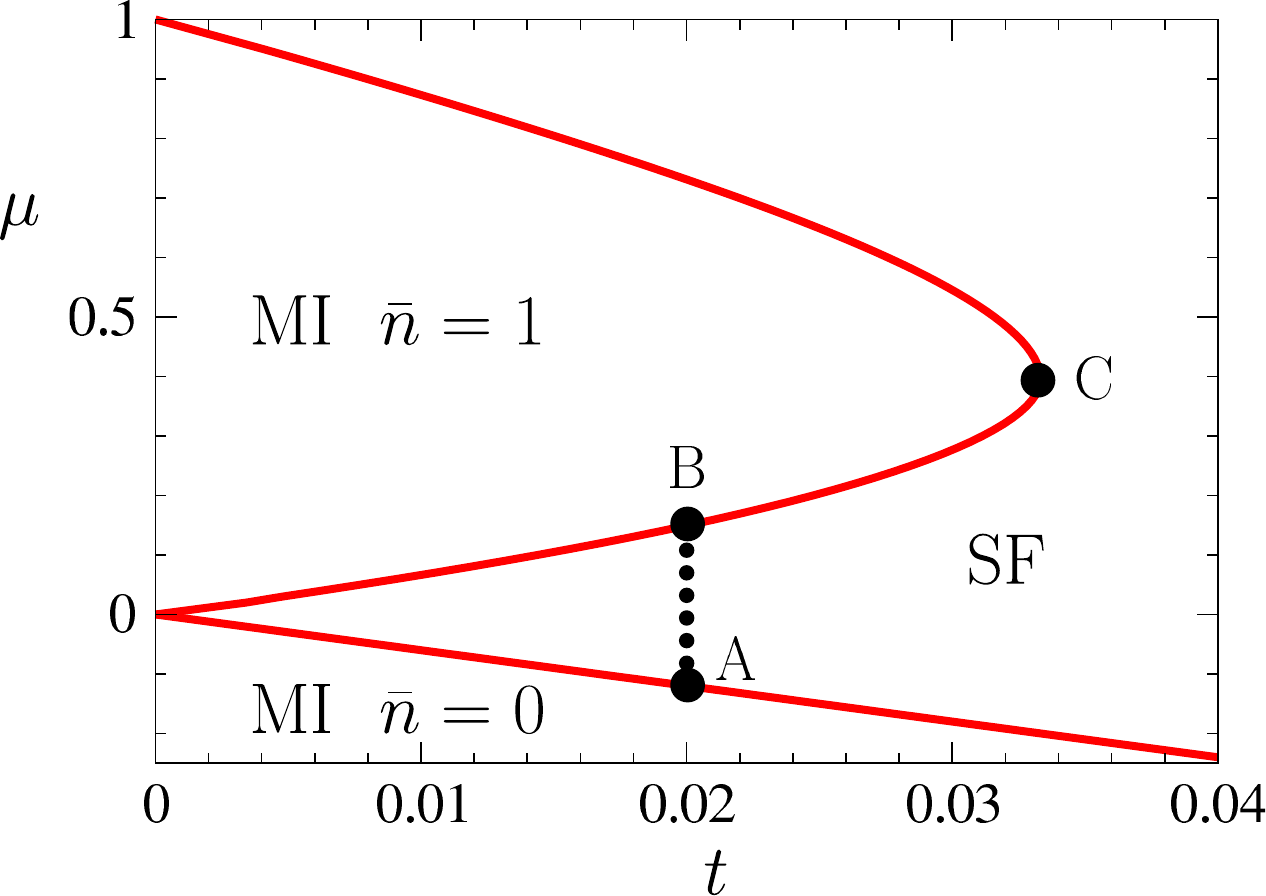}}
\vspace{-0.25cm}
\caption{(Color online) Phase diagram of the Bose-Hubbard model on a cubic lattice showing the Mott insulators (MI) with density $\bar n=0$ (vacuum) and $\bar n=1$, as well as the surrounding superfluid phase (SF). Point C at the tip of the Mott lobe shows the multicritical point where the transition occurs at fixed density $\bar n=1$. Away from this point, the transition is driven by a density change. The zero-temperature pressure $P(\mu,T=0)$ along the dotted line AB is shown in Fig.~\ref{fig_pressure}. The finite-temperature pressure $P(\mu_c,T)$ at point B is shown in Fig.~\ref{fig_pressureT}. (In all figures of the Letter, we take $U$ as the energy unit.)}
\label{fig_phase_dia}
\end{figure}

The equation of state of a dilute Bose gas is usually derived from a low-density expansion. To next-to-leading order in the parameter $ma^2\mu$, the pressure $P(\mu,T)$ depends only on the boson mass $m$ and the $s$-wave scattering length $a$ (and not on the microscopic details of atom-atom interactions). The equation of state can also be understood from the point of view of phase transitions~\cite{Sachdev_book}. By varying the chemical potential from negative to positive values at zero temperature, one induces a quantum phase transition between a state with no particles (vacuum) and a superfluid state. Elementary excitations at the QCP are free bosons of mass $m$ and the interaction between two elementary excitations is determined by the scattering length $a$ in the low-energy limit. Thus the equation of state $P(\mu,T)$ is uniquely determined by the QCP which therefore controls the thermodynamics of the dilute Bose gas.

The advantage of this point of view is that it allows us to understand the density-driven Mott transition in the Bose-Hubbard model along similar lines, using the fact that it belongs to the dilute Bose gas universality class. At the QCP between the Mott insulator and the superfluid phase, the elementary excitations are quasi-particles with effective mass $m^*$ and their mutual interaction is described by an effective scattering length $a^*$. Thus when the boson density $\bar n$ slightly differs from the commensurate density $\bar n_c$ of the Mott insulator, the excess density $|\bar n-\bar n_c|$ of particles (or holes) behaves as a dilute gas of quasi-particles~\cite{Fisher89} (with effective parameters $m^*$ and $a^*$) with an equation of state $P(\mu,T)$ which must be the same as the one of the dilute Bose gas up to some nonuniversal parameters such as $m^*$ and $a^*$. 

Our main results can be summarized as follows. Near the density-driven Mott transition and for temperature $T$ smaller than the hopping amplitude $t$, the pressure takes the form
\begin{equation}
P(\mu,T) = P_c + \bar n_c \delta\mu + {m^*}^{3/2} T^{5/2} \tilde P \left (\pm \frac{\delta\mu}{T},\pm m^*a^{*2}\delta\mu \right) ,
\label{scaling}
\end{equation}
where $P_c$ and $\bar n_c$ are the pressure and density (mean number of bosons per site) at the QCP $\mu=\mu_c$, respectively, and $\delta\mu=\mu-\mu_c$. $\tilde P(x,y)$ is a universal scaling function characteristic of the (three-dimensional) dilute Bose gas universality class. The $+$ ($-$) sign in~(\ref{scaling}) corresponds to particle (hole) doping. Equation~(\ref{scaling}) is valid everywhere near the superfluid--Mott-insulator transition except in the close vicinity of the multicritical points where the transition takes place at fixed (commensurate) density (Fig.~\ref{fig_phase_dia}). For $\mu_c=P_c=\bar n_c=0$, $m^*=m$ and $a^*=a$, Eq.~(\ref{scaling}) reproduces the low-density expansion of a dilute gas of bosons with mass $m$ and scattering length $a$. In this Letter, we discuss the equation of state~(\ref{scaling}) in two different limits.

In the zero-temperature superfluid phase, 
\begin{multline}
P(\mu,T=0) = P_c + \bar n_{c}\delta\mu \\  +\frac{m^*}{8\pi a^*}(\delta\mu)^2 \left( 1 - \frac{64}{15\pi} \sqrt{m^*a^{*2}|\delta\mu|} + \cdots\right) ,
\label{pressure}
\end{multline}
where the ellipses denote higher-order terms in the expansion parameter $m^*a^{*2}|\delta\mu|$. Taking the derivative of Eq.~(\ref{pressure}) wrt $\mu$, we obtain
\begin{equation} 
\bar n-\bar n_c = \frac{m^*\delta\mu}{4\pi a^*} \left( 1 - \frac{16}{3\pi} \sqrt{m^*a^{*2}|\delta\mu|} + \cdots\right) , 
\label{density}
\end{equation}
where $\bar n=dP/d\mu$ is the boson density. Equations~(\ref{pressure}) and (\ref{density}) are similar to the known results for a dilute Bose gas. The correction to the ``mean-field'' result $P_0=m^*(\delta\mu)^2/8\pi a^*$ is known as the Lee-Huang-Yang correction~\cite{Lee57a,*Lee57b}. 

At finite temperature and for $\mu=\mu_c$ (quantum critical regime), 
\begin{equation}
P(\mu_c,T) = P_c +\zeta(5/2) \left( \frac{m^*}{2\pi} \right)^{3/2} T^{5/2} + \cdots 
\label{pressureT}
\end{equation}
for $T\lesssim t$, where $t$ is the hopping amplitude between neighboring sites [See Eq.~(\ref{action}) below]. Once $T$ is of order of the boson dispersion $t$, the lattice starts to play a role and the universal description in terms of quasi-particles (with parameters $m^*$ and $a^*$) breaks down. Equation~(\ref{pressureT}) agrees with the well-known expression of the pressure in the dilute Bose gas when $|\mu|\ll T$~\cite{Sachdev_book}. 

{\it Equation of state.} 
To derive Eqs.~(\ref{pressure}) and (\ref{pressureT}), we start from the action of the Bose-Hubbard model, 
\begin{align}
S = \inttau \biggl\lbrace & \sum_\r \Bigl[ \psi_\r^* (\dtau-\mu)\psi_\r + \frac{U}{2} (\psi_\r^*\psi_\r)^2 \Bigr] \nonumber \\ & - t \sum_{\mean{\r,\r'}} \left(\psi_\r^* \psi_{\r'}+\mbox{c.c.}\right) \biggr\rbrace ,
\label{action}
\end{align}
where $\psi_\r(\tau)$ is a complex field and $\tau\in [0,\beta]$ an imaginary time with $\beta=1/T$ the inverse temperature. $\lbrace\r\rbrace$ denotes the $N$ sites of the lattice, $U$ the on-site repulsion, and $t$ the hopping amplitude between nearest-neighbor sites $\mean{\r,\r'}$. We set $\hbar=k_B=1$ and take the lattice spacing as the unit length throughout the Letter. 

The NPRG approach~\cite{Rancon11a,Rancon11b} allows us to compute the effective action $\Gamma[\phi^*,\phi]$ defined as the Legendre transform of the thermodynamic potential $-\ln Z[J^*,J]$. $J_\r$ is an external source which couples linearly to the bosonic field $\psi_\r$ and $\phi_\r(\tau)=\delta\ln Z[J^*,J]/\delta J^*_\r(\tau)=\mean{\psi_\r(\tau)}$ is the superfluid order parameter. Thermodynamic properties of the system can be derived from the effective potential defined by $V(n)=(\beta N)^{-1}\Gamma[\phi^*,\phi]$ with $n=|\phi|^2$ and $\phi$ a constant (uniform and time-independent) field. Its minimum determines the condensate density $n_0$ and the pressure $P=-V(n_0)$ in the equilibrium state. The single-particle propagator $G=-\Gamma^{(2)-1}$ is related to the two-point vertex $\Gamma^{(2)}$ defined as the second-order functional derivative of $\Gamma$. 

At the QCP between the superfluid phase and the Mott insulator, the effective action takes the form
\begin{align}
\Gamma[\phi^*,\phi] ={}& \inttau \int d^3r \Bigl[ \phi^*(Z_C\dtau-Z_At\nablabf^2)\phi \nonumber \\ & 
+ \frac{\lamb}{2} |\phi|^4 + \cdots \Bigr]
\label{gamma}
\end{align}
(with $\beta\to\infty$) up to a constant (field-independent) term. Since we are interested in the low-energy limit, we consider the continuum limit where $\r$ becomes a continuous variable. The ellipses denote higher-order (in derivative or field) terms. Equation~(\ref{gamma}) is valid at a generic QCP where the transition is driven by a density change. At a multicritical point, where the transition occurs at fixed (commensurate) density, $Z_C$ vanishes and one should explicitly include a $\partial_\tau^2$ term; the quantum phase transition is then in the university class of the O(2) model~\cite{Rancon11a,Rancon11b}.

\begin{figure}
\centerline{\includegraphics[width=4cm,clip]{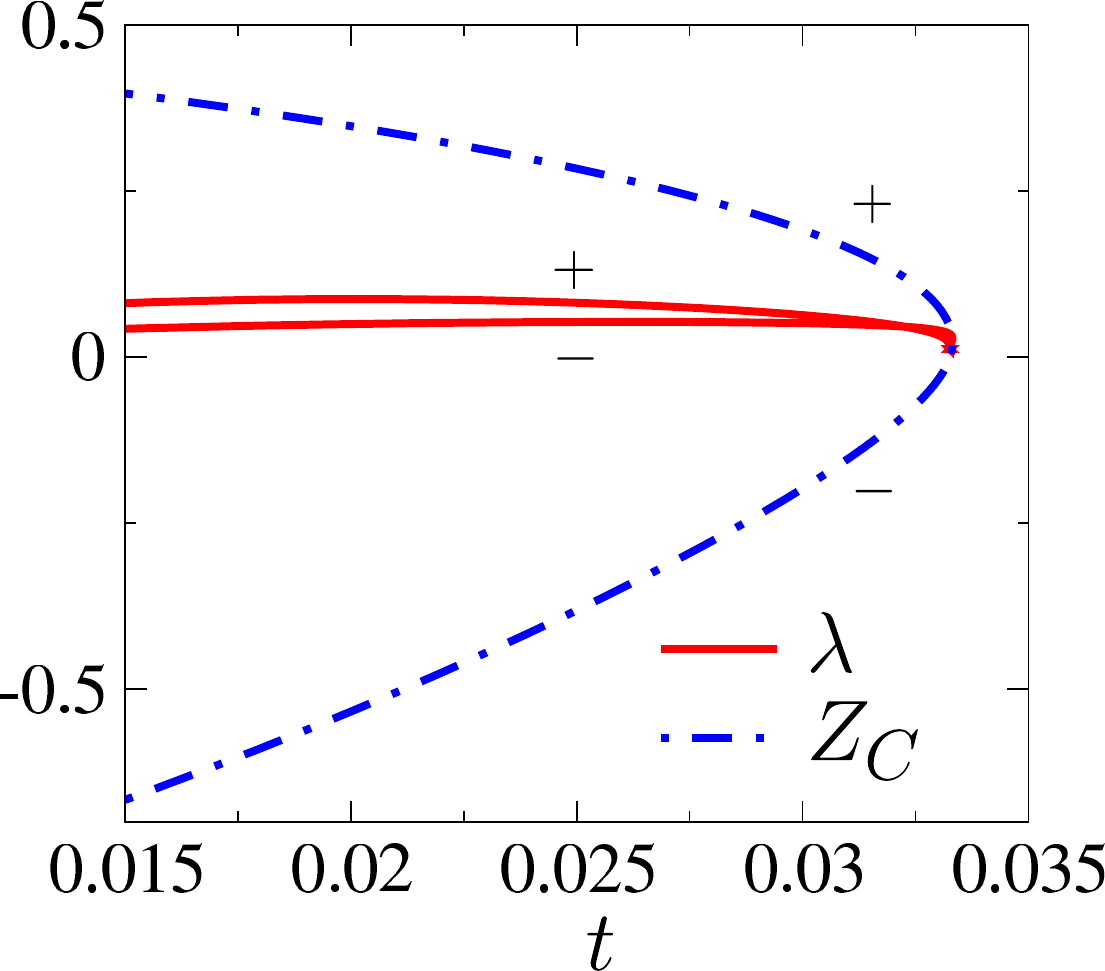}
\hspace{0.2cm}
\includegraphics[width=4cm,clip]{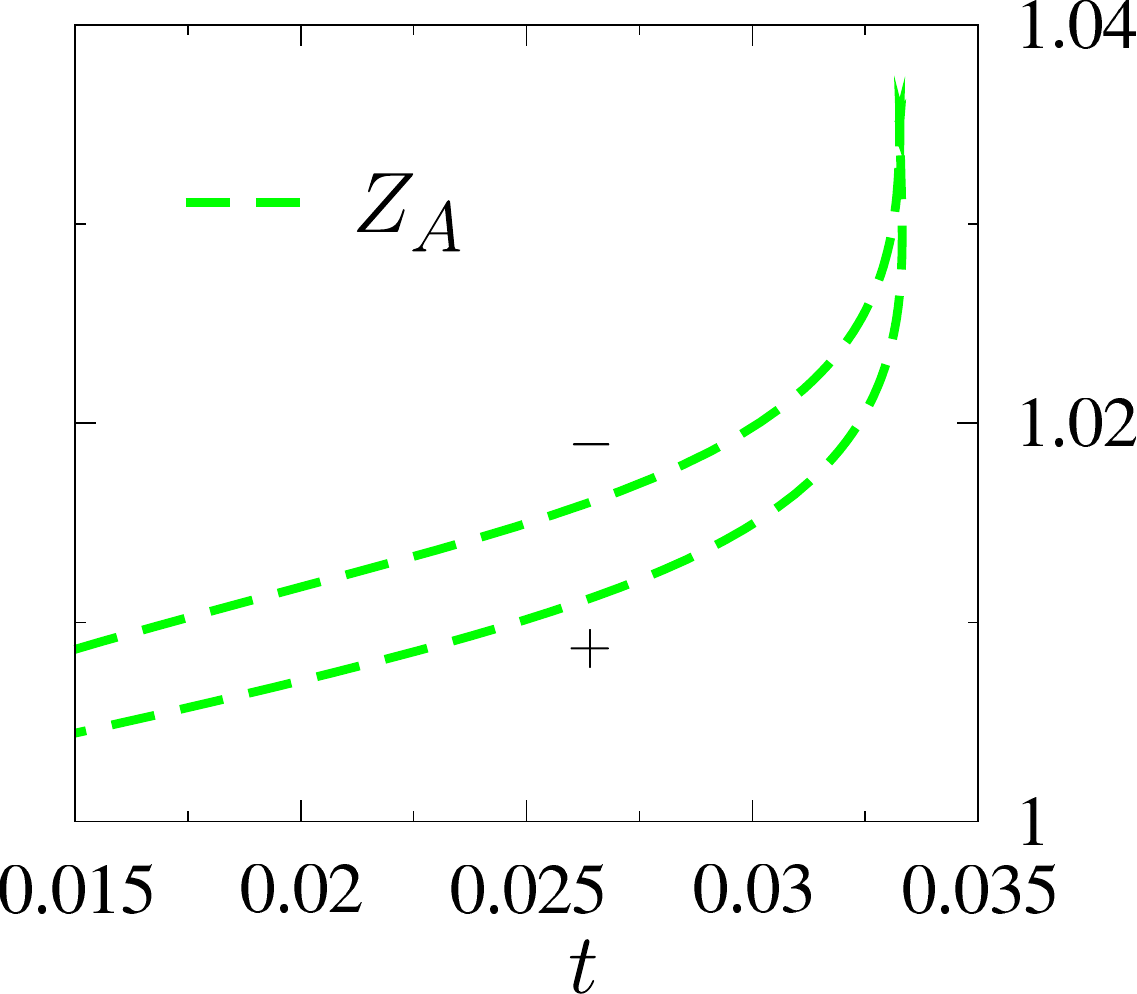}}
\vspace{-0.25cm}
\caption{(Color online) $Z_C$, $\lamb$ and $Z_A$ vs $t/U$ at the QCP between the superfluid phase and the Mott insulator $\bar n=1$. The $+$ and $-$ signs refer to the upper and lower parts of the transition line.}
\label{fig_ZCZAlamb}
\vspace{0.25cm}
\centerline{\includegraphics[width=6cm,clip]{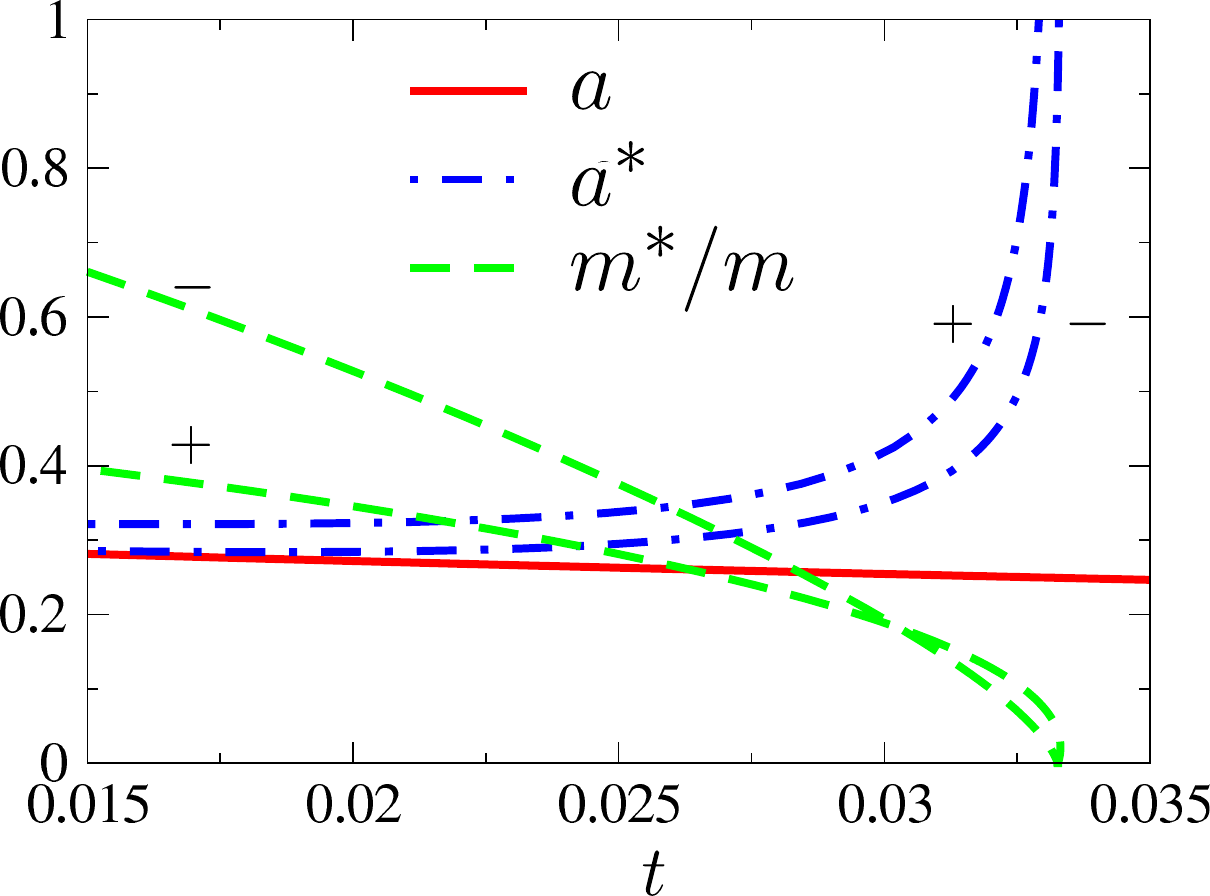}}
\vspace{-0.25cm}
\caption{(Color online) Effective mass $m^*$ and scattering length $a^*$ vs $t/U$ at the QCP  between the superfluid phase and the Mott insulator $\bar n=1$. The solid (red) line shows the scattering length $a=[8\pi(t/U+A)]^{-1}$ (see text).}
\label{fig_ma_eff}
\end{figure}

From Eq.~(\ref{gamma}), we can identify the elementary excitations at the QCP. $Z_C$ is negative on the lower part of the transition line (for a given Mott lobe). In that case, it is convenient to perform a particle-hole transformation $\phi\leftrightarrow\phi^*$ (which changes the sign of the $\dtau$ term in~(\ref{gamma})). We can then define a quasi-particle field $\bar\phi=|Z_C|^{1/2}\phi$ and rewrite the effective action as 
\begin{align}
\Gamma[\bar\phi^*,\bar\phi] ={}&  \inttau \int d^3r \Bigl[ \bar\phi^*\Bigl(\dtau-\frac{\nablabf^2}{2m^*} \Bigr)\bar\phi \nonumber \\ & + \half \frac{4\pi a^*}{m^*} |\bar\phi|^4 + \cdots \Bigr] ,
\label{gamma1}
\end{align} 
where 
\begin{equation}
m^* = \frac{|Z_C|}{2tZ_A} = m \frac{|Z_C|}{Z_A} , \quad
a^* = \frac{m^*\lamb}{4\pi Z_C^2} , 
\label{maeff}
\end{equation}
with $m=1/2t$ the effective mass of the free bosons moving on the lattice. The elementary excitations are quasi-particles with a quadratic dispersion law and mass $m^*$. They are particle-like if $Z_C>0$ and hole-like if $Z_C<0$, and the quasi-particle weight is $|Z_C|^{-1}$. The interaction between two quasi-particles is determined by the effective scattering length $a^*$. 

At the QCP between the superfluid phase and the vacuum (the Mott insulator with vanishing density), $Z_C=Z_A=1$ (the single-particle propagator is not renormalized~\cite{Rancon11b}) so that $m^*=m=1/2t$. Furthermore, the interaction constant $\lamb=8\pi ta$ can be calculated analytically and related to the scattering length $a=[8\pi(t/U+A)]^{-1}$ ($A\simeq 0.1264$) of the bosons moving on the lattice~\cite{Rancon11b}, which gives $a^*=a$. For a generic QCP between the superfluid phase and a Mott phase with nonzero density ($\bar n_c=1,2,\cdots$), the determination of $m^*$ and $a^*$ is much more difficult as it requires to solve a many-body problem. $Z_C$, $Z_A$ and $\lamb$ can be  obtained from the numerical solution of the NPRG equations~\cite{Rancon11b}. Figures~\ref{fig_ZCZAlamb} and \ref{fig_ma_eff} show $Z_C$, $Z_A$, $\lamb$, $m^*$ and $a^*$ as a function of $t/U$ for the QCP separating the superfluid phase from the Mott insulator $\bar n=1$. The vanishing of $Z_C$ at the multicritical point implies that $m^*$ vanishes and $a^*$ diverges when we approach the tip of the Mott lobe (point C in Fig.~\ref{fig_phase_dia}). Note that it is numerically difficult to determine $m^*$ and $a^*$ for $t/U\lesssim 0.015$ due to the degeneracy between the states with density $\bar n=p$ and $\bar n=p+1$ when $\mu/U=p$ ($p$ integer) and $t=0$. 

\begin{figure}
\centerline{\includegraphics[width=5.5cm]{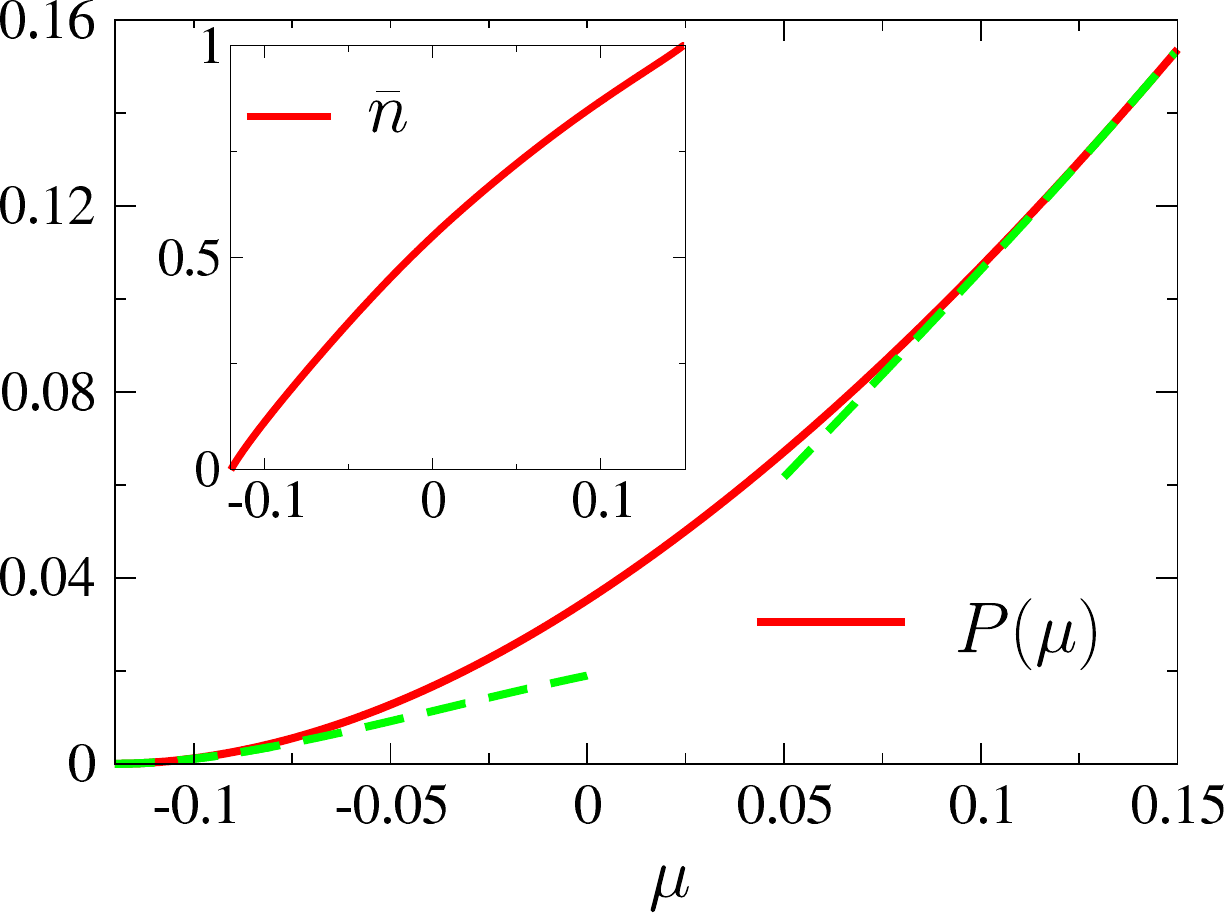}}
\vspace{0.25cm}
\centerline{\includegraphics[width=3.9cm]{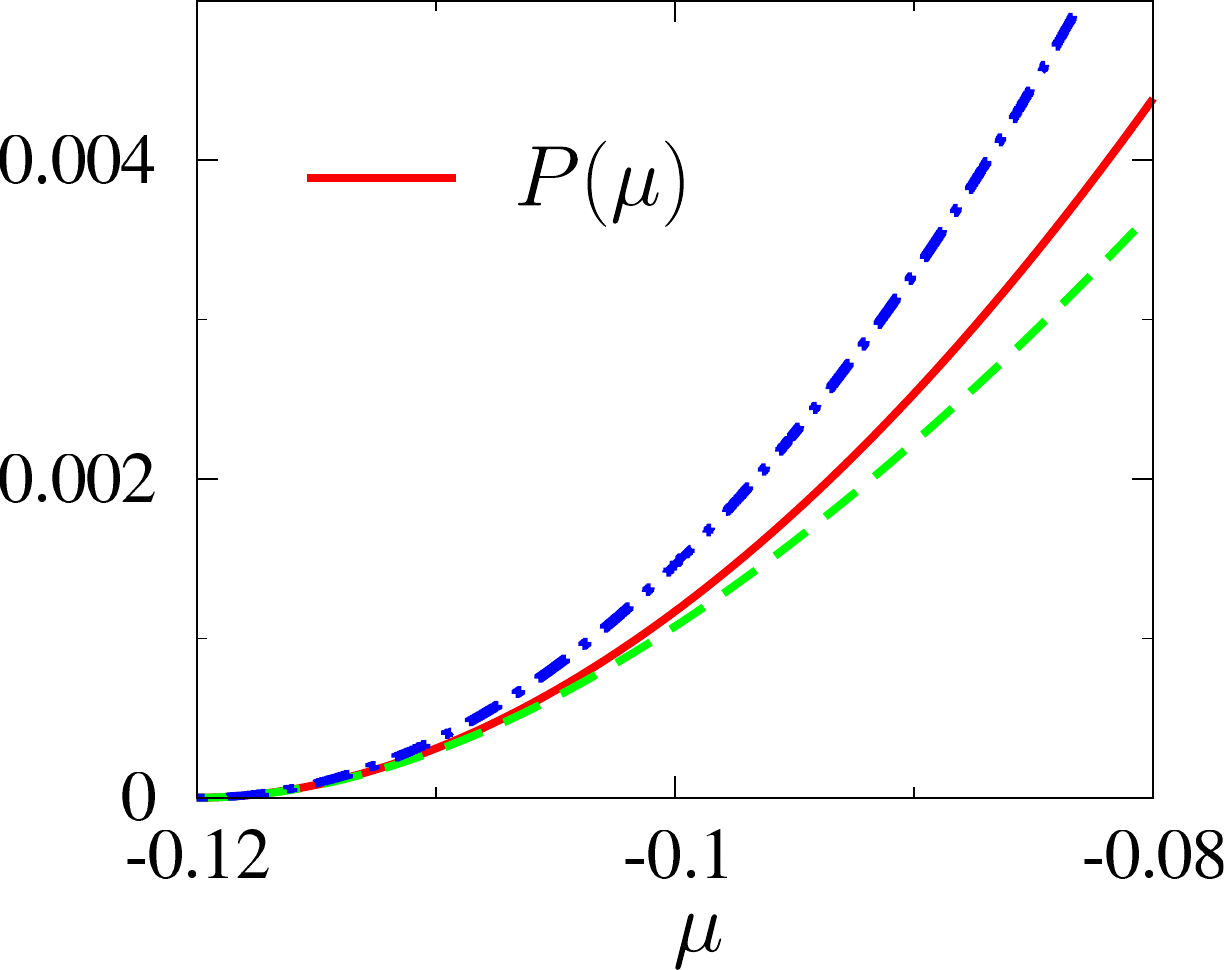}
\includegraphics[width=4.1cm]{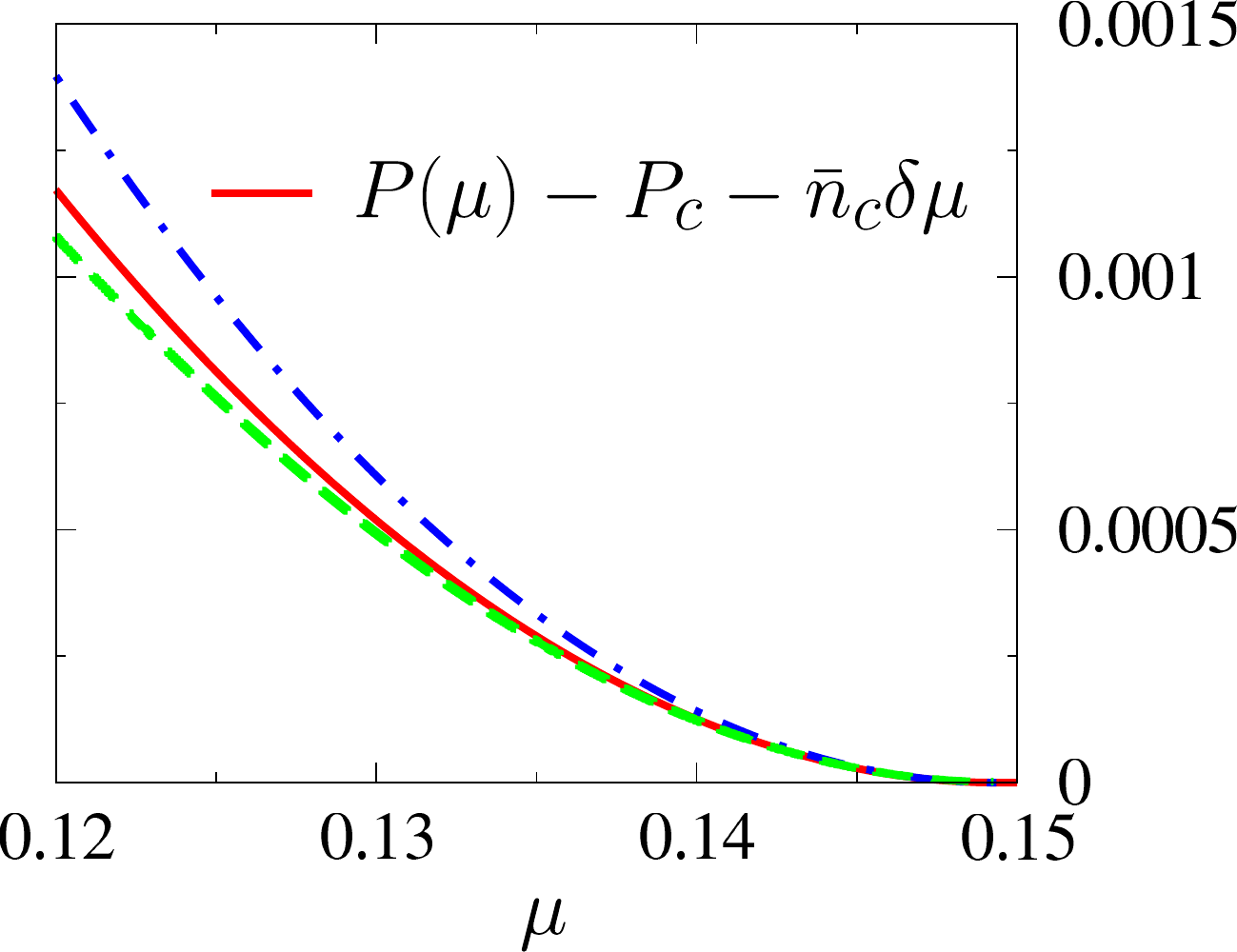}}
\vspace{-0.25cm}
\caption{(Color online) Pressure $P$ vs chemical potential $\mu$ along the dotted line in Fig.~\ref{fig_phase_dia}, as obtained from the NPRG approach (the inset shows the density $\bar n=dP/d\mu$). The bottom figures show the behavior near the Mott insulating phases $\bar n=0$ and $\bar n=1$. The dashed (green) line corresponds to Eq.~(\ref{pressure}) and the dash-dotted (blue) one to the ``mean-field'' result $P=P_c+\bar n_c\delta\mu+(\delta\mu^2)m^*/8\pi a^*$.}
\label{fig_pressure}
\end{figure}

We are now in a position to verify that near the QCP the zero-temperature pressure in the superfluid phase is given by Eq.~(\ref{pressure})~\cite{note03}. Note that higher-order terms neglected in Eq.~(\ref{gamma}) do not contribute to the pressure to the order considered~\cite{Braaten01}.  
Figure~\ref{fig_pressure} shows $P(\mu,T=0)$ at fixed $t/U$ and for a density varying between 0 and 1 (see the dotted line AB in Fig.~\ref{fig_phase_dia}). Near the Mott insulating phases $\bar n=0$ and $\bar n=1$, we find a very good agreement between the NPRG result and Eq.~(\ref{pressure}). As we approach the tip of the Mott lobe located at ($\bar t_c,\bar\mu_c$), $m^*a^{*2}$ diverges and the domain of validity $|\delta\mu|\ll 1/m^*a^{*2}$ of Eq.~(\ref{pressure}) shrinks to zero. For $t=\bar t_c$ and $\mu\simeq \bar\mu_c$, the pressure varies as $P(\mu,T=0)\simeq P_c+\bar n_c \delta\mu+\alpha(\mu-\bar\mu_c)^4$ with $\alpha$ a constant.

\begin{figure}
\centerline{\includegraphics[width=6cm]{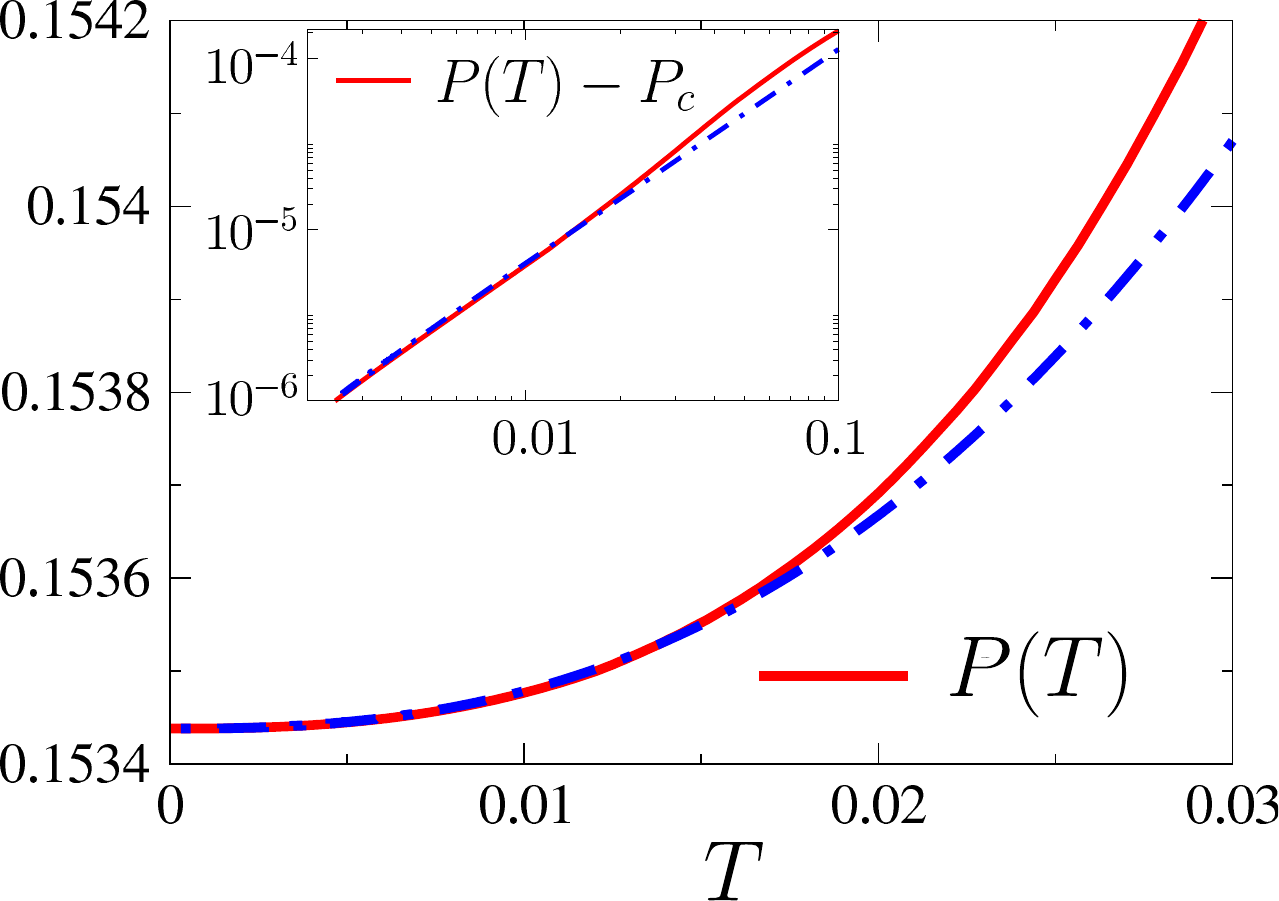}}
\vspace{-0.25cm}
\caption{(Color online) Pressure $P(\mu_c,T)$ vs temperature $T$ [$t=0.02U$]. The dash-dotted (blue) line corresponds to Eq.~(\ref{pressureT}). The inset shows a log-log plot and the $T^{5/2}$ dependence of $P(\mu_c,T)-P_c$.}
\label{fig_pressureT}
\end{figure}

The finite-temperature pressure $P(\mu_c,T)$ at point B in Fig.~\ref{fig_phase_dia} is shown in Fig.~\ref{fig_pressureT}. We obtain a very good agreement with Eq.~(\ref{pressureT}) as long as lattice effects can be ignored, i.e. $T\lesssim t$. One can also compute the temperature dependence of the pressure for $\mu\neq \mu_c$. In all cases, one recovers the known results of the dilute Bose gas~\cite{Sachdev_book} but with $m$ and $a$ replaced by $m^*$ and $a^*$.

{\it Conclusion.} We have shown that the pressure $P(\mu,T)$ of a Bose gas near the density-driven Mott transition takes a universal form, analog to that of a dilute Bose gas, once a certain set of low-energy parameters have been fixed. These parameters characterize the elementary excitations at the QCP and their interactions, and can be systematically computed using the NPRG approach. In three-dimensions (i.e. above the upper critical dimension $d_c=2$ of the $T=0$ superfluid--Mott-insulator transition), this set is formally infinite but to leading order and for $T\ll t$ only the mass $m^*$ and scattering length $a^*$ of the elementary excitations is needed [Eq.~(\ref{scaling})]. At the upper critical dimension $d_c=2$, our analysis still holds; universality is even stronger, as only $m^*$ and $a^*$ need to be known (to all orders in the small parameter $m^*a^{*2}|\delta\mu|$ and for $T\ll t$) to determine the equation of state. 

Recent experiments in cold atom gases have shown that it is now possible to deduce the pressure $P(\mu,T)$ of an homogeneous infinite system from the doubly-integrated {\it in situ} density profile $\int dxdy\,\bar n(x,y,z)$ of an harmonically trapped gas~\cite{Cheng07,*Shin08,*Ho09,*Nascimbene10,*Navon10}. The zero-temperature equation of state (including the Lee-Huang-Yang correction) of a homogeneous Bose gas of $^7$Li atoms has been measured using this technique~\cite{Navon11}. Experiments with atoms loaded in an optical lattice~\cite{Trotzky10} are now approaching the low-temperature regime where our predictions for $P(\mu,T=0)$ [Eq.~(\ref{pressure})] could be observed. Recently, quantum criticality of a two-dimensional Bose gas near the vacuum-superfluid quantum phase transition has been observed at finite temperature, thus explicitly demonstrating that the equation of state of cold atomic gases gives direct information about the QCP~\cite{Zhang11}. We expect that similar measurements will allow to observe quantum criticality near the (nontrivial) QCP governing the superfluid--Mott-insulator transition, and verify our predictions for the effective mass $m^*$ and scattering length $a^*$ in three-dimensional Bose gases.   

We would like to thank X. Leyronas, F. Chevy and C. Salomon for discussions. 

\vspace{-0.75cm}

%


\begin{thebibliography}{28}
\bibitem [{not()}]{note03}%
  \BibitemOpen
  \href@noop {} {}\bibinfo {note} {The compressibility $\kappa=d\bar
  n/d\mu=d^2P/d\mu^2$ (which has no regular part at the
  superfluid--Mott-insulator transition since it vanishes in the Mott
  insulating phase) takes a form similar to that of a dilute Bose gas. By
  integrating $\kappa$ wrt $\mu$, one obtains
  Eqs.~(\ref{pressure},\ref{density}).}\BibitemShut {Stop}%
\end{thebibliography}


\end{document}